\documentclass[sigconf,nonacm,9pt]{acmart}

\AtBeginDocument{%
  \providecommand\BibTeX{{%
    Bib\TeX}}}

\setcopyright{acmlicensed}
\copyrightyear{2018}
\acmYear{2018}
\acmDOI{XXXXXXX.XXXXXXX}

\acmJournal{JDS}
\acmVolume{37}
\acmNumber{4}
\acmArticle{111}
\acmMonth{8}

\usepackage{xcolor}
\usepackage{listings}
\usepackage{tikz}
\usepackage{amsmath}
\usepackage{multirow}
\usepackage{tabularx}

\usepackage{setspace}
\usepackage{enumitem}
\setlist[itemize]{noitemsep, topsep=0pt, leftmargin=*}

\DeclareRobustCommand\encircle[1]{\tikz[baseline=(char.base)]{
  \node[shape=circle, fill,        
        inner sep=0pt,              
        minimum size=2.6ex]           
        (char) {\textcolor{white}{#1}}; 
}}

\usepackage{graphicx}
\def\BibTeX{{\rm B\kern-.05em{\sc i\kern-.025em b}\kern-.08em
    T\kern-.1667em\lower.7ex\hbox{E}\kern-.125emX}}

\lstdefinestyle{sqlstyle}{
  language=SQL,
  basicstyle=\ttfamily\small,
  keywordstyle=\color{blue}\bfseries,
  commentstyle=\color{teal!60},
  stringstyle=\color{orange!70!black},
  numbers=none,
  showstringspaces=false,
  columns=fullflexible,
  breaklines=true
}

\begin{document}

\title{HDDB: Efficient In-Storage SQL Database Search Using Hyperdimensional Computing on Ferroelectric NAND Flash}


\author{
  Quanling Zhao\textsuperscript{*}$^{1}$, Yanru Chen\textsuperscript{*}$^{1}$, Runyang Tian$^{1}$, Sumukh Pinge$^{1}$, Weihong Xu$^{2}$, Augusto Vega$^{3}$, Steven Holmes$^{3}$, Saransh Gupta$^{3}$, Tajana Rosing$^{1}$\\
  $^{1}$ UCSD, $^{2}$ EPFL, $^{3}$ IBM\\
  }
\email{{quzhao,yac054,r3tian,spinge,tajana}@ucsd.edu}
\email{weihong.xu@epfl.ch}
\email{ajvega@us.ibm.com, {Steven.Holmes,saransh}@ibm.com}

\thanks{\textsuperscript{*}Both authors contributed equally to this research.}

\acmArticleType{Review}
\acmCodeLink{https://github.com/borisveytsman/acmart}
\acmDataLink{htps://zenodo.org/link}
\acmContributions{BT and GKMT designed the study; LT, VB, and AP
  conducted the experiments, BR, HC, CP and JS analyzed the results,
  JPK developed analytical predictions, all authors participated in
  writing the manuscript.}


\begin{abstract}

Hyperdimensional Computing (HDC) encodes information and data into high-dimensional distributed vectors that can be manipulated using simple bitwise operations and similarity searches, offering parallelism, low-precision hardware friendliness, and strong robustness to noise. These properties are a natural fit for SQL database workloads dominated by predicate evaluation and scans, which demand low energy and low latency over large fact tables. Notably, HDC’s noise-tolerance maps well onto emerging ferroelectric NAND (FeNAND) memories, which provide ultra-high density and in-storage compute capability but suffer from elevated raw bit-error rates. In this work, we propose HDDB, a hardware–software co-design that combines HDC with FeNAND multi-level cells (MLC) to perform in-storage SQL predicate evaluation and analytics with massive parallelism and minimal data movement. Particularly, we introduce novel HDC encoding techniques for standard SQL data tables and formulate predicate-based filtering and aggregation as highly efficient HDC operations that can happen in-storage. By exploiting the intrinsic redundancy of HDC, HDDB maintains correct predicate and decode outcomes under substantial device noise (up to 10\% randomly corrupted TLC cells) without explicit error-correction overheads. Experiments on TPC-DS fact tables show that HDDB achieves up to 80.6× lower latency and 12,636× lower energy consumption compared to conventional CPU/GPU SQL database engines, suggesting that HDDB provides a practical substrate for noise-robust, memory-centric database processing.

\end{abstract}

\maketitle

\section{Introduction}

Predicate evaluation—evaluating boolean conditions over rows of a data table—is the front door of almost every SQL database query: it determines which data flow to downstream operators and thus often dominates end-to-end cost~\cite{zou2020empowering}. In workloads over large fact tables, these predicates are typically applied as scans; the cost is dominated by data movement rather than computation. All values must be read, compared, and filtered, saturating memory bandwidth and wasting energy shuttling bytes that will mostly be discarded. This has motivated a long line of near-data and in-storage processing proposals that offload filtering into or near storage devices to reduce host–storage traffic for data-intensive queries~\cite{kim2011fast,kim2016storage,jo2016yoursql,bernhardt2023pimdb,kim2024darwin,lu2024sql2fpga}.

Hyperdimensional Computing (HDC) provides a particularly suitable algorithmic foundation for such workloads. HDC represents symbols and data as high-dimensional, often low-precision (e.g., binary) vectors—hypervectors (HVs)—and manipulates them using simple bitwise operations and similarity search~\cite{kanerva2009hyperdimensional,chung2025robust,thomas2021theoretical,ge2020classification}. These operations map naturally to wide, bit-parallel datapaths and can be implemented close to where data resides, making HDC an attractive fit for in- or near-storage processing of scan-heavy SQL predicates. As a result, HDC can deliver low-latency, low-energy evaluation of simple decisions over large SQL fact tables, while its distributed representations also offer inherent robustness to device-level noise. Recent work has mapped HDC primitives onto emerging memory and logic devices, but these efforts have largely targeted AI or specialized database (e.g., Mass spectrometry) workloads rather than general SQL database workloads~\cite{karunaratne2020memory,thomann2022all,kang2025relhdx,dutta2022hdnn,pinge2025fenoms}.

Emerging ferroelectric NAND (FeNAND) flash offers an appealing substrate for HDC-based SQL predicate processing. By leveraging ferroelectrics in CMOS-compatible processes, FeNAND promises ultra-high density, low power, and high-speed, three-dimensional integration—properties that are attractive for database storage and in-storage compute capabilities~\cite{kim2021ferroelectric}. However, FeNAND’s narrow sense margins and retention-induced shifts introduce non-negligible readout noise and bit errors, which can severely strain conventional ECC-centric designs. In this context, HDC’s parallel, noise-robust computation is a natural match for noisy but highly efficient FeNAND arrays, and predicate scans in SQL databases are exactly the kind of simple, massively parallel decisions that HDC accelerates well.

In this work, we bring these threads together and propose HDDB, a hardware–software co-design that combines HDC and FeNAND for fast and energy-efficient in-storage SQL predicate evaluation and analytics with massive parallelism and minimal data movement. Our contributions are as follows:

\textbf{(1)} HDDB introduces, to the best of our knowledge, the first method for encoding conventional SQL tables into binary HDC representations while preserving decodability back to the original information to ensure that the data can be read from the hypervectors. On top of these encodings, HDDB provides predicate evaluation algorithms for both string and numerical conditions using only standard HDC primitives—bitwise operations and Hamming-distance-based similarity search—making it the first system to apply HDC to SQL database workloads and providing a general framework that can be extended to broader SQL CRUD operations.

\textbf{(2)} On the hardware side of HDDB, we propose a mapping from binary HDC representations to FeNAND triple-level cell (TLC) that packs multiple bits into each cell. This enables exploiting FeNAND’s intrinsic advantages to execute HDC operations where the data resides. We further propose an In-Storage-Processing (ISP) accelerator. Integrating 3D FeNAND dies with near-storage processors (NSPs) to minimize data movement, our architecture is engineered to efficiently execute predicate-based filtering and additional analytics such as common aggregation functions. Together, the HDC encoding, predicate algorithms, and ISP design convert FeNAND MLC from a noisy storage medium into an efficient substrate for robust, energy-efficient processing platform for SQL databases.

\textbf{(3)} We evaluate HDDB on TPC-DS~\cite{TPC-DS} fact tables across multiple scaling factors, comparing against state-of-the-art (SOTA) SQL database engines and in-/near-storage baselines. Our results show that HDDB maintains correct predicate outcomes under substantial device noise (e.g., up to 10\% randomly corrupted MLC cells) while running up to $80.6\times$ faster and consuming $12{,}636\times$ less energy.

\section{Background and Related Works}

\textbf{Hyperdimensional Computing (HDC): }HDC is an approach to representing symbols, sets, and relations with random, high-dimensional, and often low-precision codes (e.g., $10^3$–$10^5$ binary bits), manipulated with a small set of algebraic operations~\cite{schlegel2022comparison,kleyko2022survey}. HDC has been applied to a wide range of applications, such as machine learning, associative memory, and many more~\cite{zhao2025bridging,dewulf2024hyperdimensional,neubert2019introduction,imani2019binary}. Previous works have shown HDC's ability to encode different structures into vectors with information retrieval capability, such as graphs, ordered sets, or even functions~\cite{raviv2024linear,poduval2022graphd,yuan2023decodable}. Since information in HDC representation is distributed across many dimensions, small fractions of bit flips rarely affect the quality of results, yielding excellent performance in noisy conditions. This “redundancy by design”~\cite{thomas2021theoretical} contrasts with error-correcting codes that treat reliability separately from computation. In this work, we provide the first system that applies HDC to SQL-style database predicate evaluation and analytics.

\noindent\textbf{Ferroelectric NAND and In-Storage Processing: }As databases reach multi-terabyte and petabyte scales, the dominant cost in analytics is moving data from storage to compute~\cite{imani2018nvquery}. ReRAM/PCM-based PIM approaches attack this bandwidth wall but lag 3D NAND in density and maturity~\cite{long2018reram,mittal2018survey}. FeNAND instead offers ultra-high-density 3D NAND suitable for multi-terabyte fact tables~\cite{kim2021ferroelectric}, but its serial string structure, narrow sense margins, and retention drift raise raw bit-error rates~\cite{cai2015read}. These characteristics call for digital, page-parallel in-storage processing that can tolerate bit errors while exploiting FeNAND’s density—the regime that HDDB targets.

\noindent\textbf{Hardware Database Accelerators: }Beyond CPU and GPU database engines, prior work accelerates database queries using ASICs, FPGAs, and computational storage: Q100-style processors implement queries in custom datapaths~\cite{wu2014q100}, FPGA systems compile SQL to reconfigurable fabrics~\cite{papaphilippou2018accelerating,fang2020memory}, and SmartSSD-like devices attach FPGAs or embedded cores to SSDs~\cite{fakhry2023review}. These designs generally assume error-free memories and focus on mapping operators onto reliable DRAM, HBM, or flash. HDDB takes a different approach: rather than adding an accelerator beside conventional NAND, it co-designs HDC encodings and FeNAND-resident in-storage processing so that the noisy flash array itself becomes the predicate engine. This cuts data movement and error-correction overhead, making predicate processing faster and more energy-efficient.

\begin{figure}[t]
    \centering
    \includegraphics[width=0.8\linewidth]{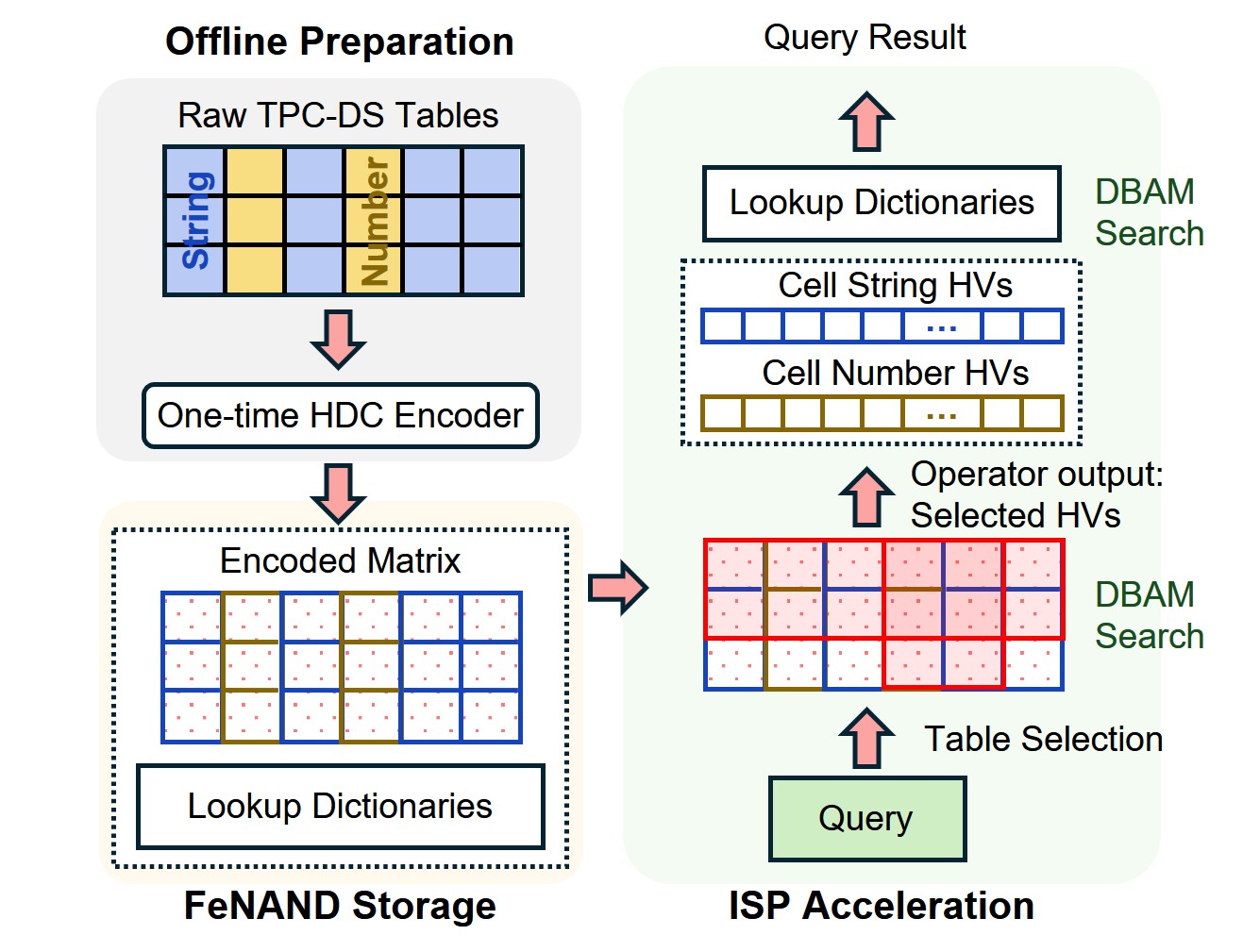}
    \caption{End-to-end HDDB predicate evaluation pipeline and acceleration targets}
    \label{fig:target}
\end{figure}

\section{HDDB Algorithm Design}
\label{sec:algorithm}

In SQL, predicate evaluation is invoked whenever the system evaluates Boolean conditions in clauses such as WHERE, JOIN ON, and HAVING, with WHERE-based filtering being the most common case. Because almost every analytic query includes one or more such predicates, and they are often applied as full-table scans, predicate evaluation frequently becomes the main bottleneck in both latency and energy~\cite{imani2018nvquery,idreos2012monetdb}. Conceptually, such queries reduce to evaluating a Boolean predicate on each row of a table to produce a selection mask, then projecting the required columns for the rows marked true. In practice, this is dominated by scanning: reading large volumes of data, comparing values against constants or ranges, forming a mask, and forwarding only selected rows. The overall dataflow of HDDB is shown in Figure~\ref{fig:target}. In HDDB, we represent data tables as long binary HVs and formulate common predicate evaluations as HDC primitives that can be executed directly in FeNAND storage.

\textbf{HDC Primitives: }We work in the standard binary HDC setting where an HV is a long bitstring. For a large dimension, independently sampled random HVs are nearly orthogonal: their Hamming distance concentrates, so flipping a modest fraction of bits barely changes relative similarity and yields noise-robustness and capacity. We use three classic HDC primitives. \emph{Bind} combines two HVs with elementwise XOR ($\otimes$), effectively “tagging’’ a value with a key and supporting approximate unbinding via $(a \otimes b) \otimes a \approx b$. \emph{Bundle} ($\oplus$) superposes multiple HVs by component-wise majority, forming a set-like representation that preserves membership in a distributed way. Finally, \emph{associative recall} compares a (possibly noisy) query HV against a dictionary using Hamming similarity and returns the nearest match. Together, these operations let us encode, store, and access SQL tables in the HDC domain.

\subsection{Mapping SQL Table into Hypervectors}

\begin{figure}[t]
    \centering
    \includegraphics[width=0.8\linewidth]{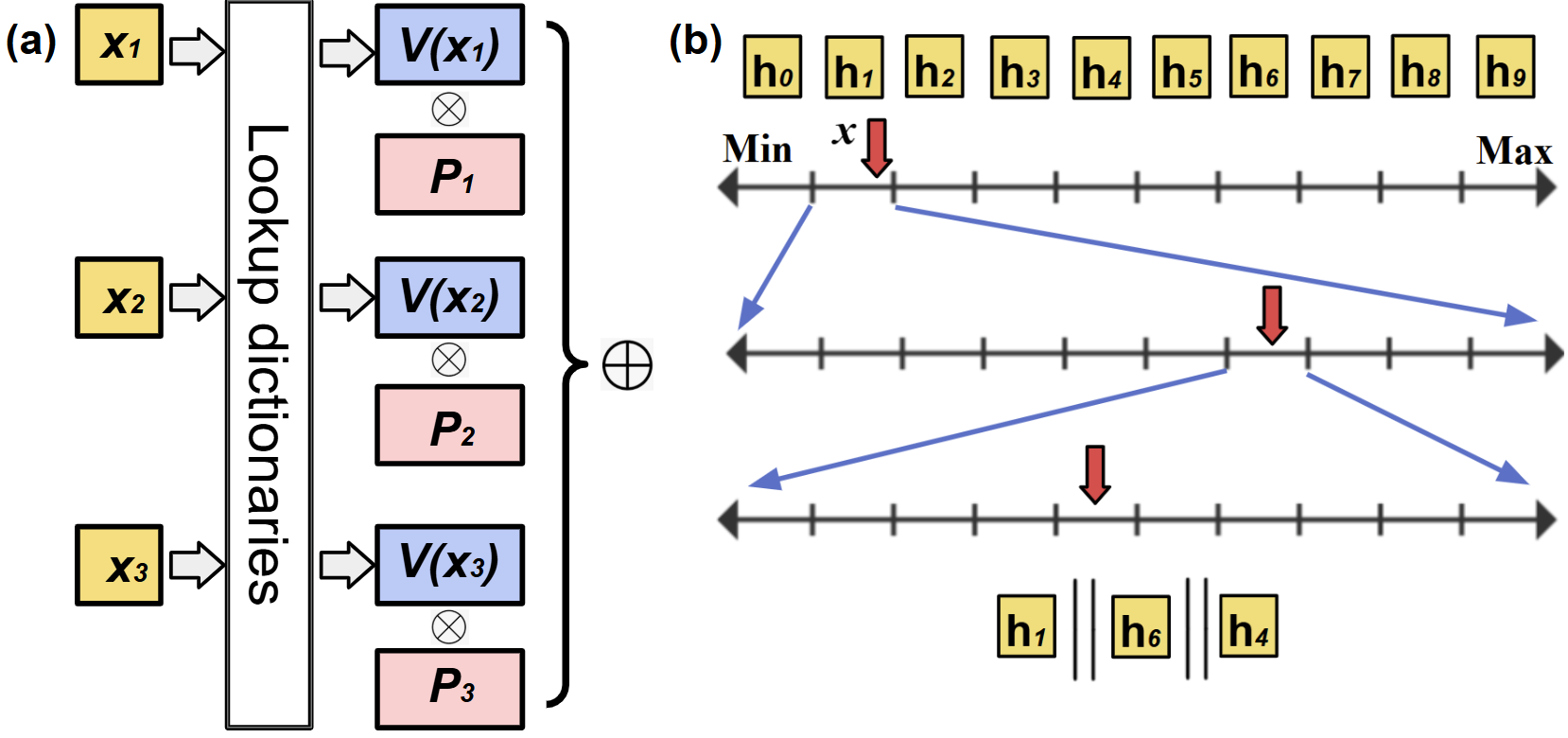}
    \caption{Mapping (a) String data (b) Numeric data into HV}
    \label{fig:encde}
\end{figure}

In SQL tables, every column behaves as either categorical or ordered data, which maps to “string-like” vs “numeric-like” for predicate semantics. For that reason, we propose different HDC encoding methods for string data and numerical data as shown in Fig~\ref{fig:encde}. To encode a string $x=x_1 x_2 \cdots x_L$ over an alphabet $\Sigma$ into a single binary HV, we first assign each symbol an i.i.d. random code $V: \Sigma \rightarrow\{0,1\}^D$ and each position a (near-orthogonal) positional HV $P_i \in\{0,1\}^D$. Each character is bound to its position, and the string HV is the bundle: $S(x)=\oplus\left(V\left(x_1\right) \otimes P_1, \ldots, V\left(x_L\right) \otimes P_L\right)$. This forms a decodable structure where the symbol at any position can be recovered with high probability via: 
\begin{equation}
\label{eq:1}
{x_i}=\arg \min _{s \in \Sigma} \operatorname{hamming}\left(S(x) \otimes P_i, V(s)\right)
\end{equation}
Intuitively, after bundling, the probe for position $i$ ($S(x) \otimes P_i$) is closer to the true symbol HV $V\left(x_i\right)$ than any other symbols by a positive similarity margin. Since the string can vary in length, we use a special termination symbol to indicate the end of a string.

In SQL databases, numeric columns behave differently from strings because their domain is ordered and often wide, so we need encodings that preserve order and make range comparisons cheap. The usual HDC approach cuts the number line into bins and each bin gets a random HV. This works coarsely but breaks down at database scale: to avoid false merges across a large span, millions of bins are needed, which explodes dictionary size and makes numerical predicates expensive.

We encode a numeric value $x \in[a, b)$ with a recursive multi-resolution scheme that reuses a small dictionary of $m$ bin HVs across $n$ levels. At level 1 we map $x$ to its coarse bin index $j_1(x)= \left\lfloor m \frac{x-a}{b-a}\right\rfloor$; at each deeper level $\ell>1$ we subdivide only the previously chosen bin and take $j_{\ell}(x)= \left\lfloor m \frac{x-x^{(\ell-1)}}{(b-a) / m^{\ell-1}}\right\rfloor$ with $x^{(\ell)}=x^{(\ell-1)}+\frac{b-a}{m^{\ell}} j_{\ell}(x)$. Each digit $j_{\ell}(x) \in\{0, \ldots, m-1\}$ selects one HV from a shared dictionary $\left\{\mathbf{h}_0, \ldots, \mathbf{h}_{m-1}\right\}$, and the final encoding is the concatenation of level HVs $E_{\text {num }}(x)=\left[\mathbf{h}_{j_1(x)}\left\|\mathbf{h}_{j_2(x)}\right\| \cdots \| \mathbf{h}_{j_n(x)}\right]$. This realizes $m^n$ distinct finest bins using only $m$ HVs and $n$ segments (memory $O(m D)$ vs. $O\left(m^n D\right)$ for naïve HDC method). This yields order-preserving numerical encoding that supports comparisons with simple HDC primitives, decodes cleanly, and is far more memory- and compute-efficient than a naïve HDC scheme that materializes one HV per fine bin.

\subsection{Predicate Evaluation \& Decode} For string predicates, the string encoder plays a hash-like role: it maps a string to an HV such that identical strings map to the same (or highly similar if under noise) HV, whereas different strings land near the expected hamming distance between two random HVs. Effectively, for exact string matching predicates, the decision is whether a row’s HV is sufficiently similar to the query HV (with a predefined threshold), a check realized entirely with HDC primitives—so false matches are increasingly unlikely with larger dimension, and predicate evaluation results emerge without decoding. To evaluate a numerical predicate, we perform associative recall over the bin dictionary per level: for each segment of the stored numeric code, run Hamming nearest-neighbor against the $m$ bin HVs to recover the level’s bin index. The recovered index sequence is then compared to the query’s index sequence. Both string and numeric predicates are inherently noise-robust in our HDC design. Bit flips on encoded tables only erode similarity search slightly, but not decisions, so predicate evaluation remains unaffected. This robustness is a direct consequence of distributed, high-dimensional HVs. After obtaining the predicate results, selected HVs can be decoded back to the original information via associative recall with the symbol dictionary.


\section{HDDB In-Storage Accelerator}

In this section, we detail the hardware–software co-design of the HDDB in-storage accelerator. This architecture is built to efficiently execute the HDC predicate evaluation algorithms presented in Section \ref{sec:algorithm} and supports additional analytics such as common aggregation functions \textbf{\{COUNT,SUM,AVG,MIN,MAX\}}. We first introduce the core organization and the overall system dataflow. We then describe the Dual Boundary Approximate Matching (DBAM) in-situ approximate searching technique, the design of specialized peripheral processors, and the column-wise data mapping strategy.

\subsection{System Overview}

\begin{figure}[t]
    \centering
    \includegraphics[width=1\linewidth]{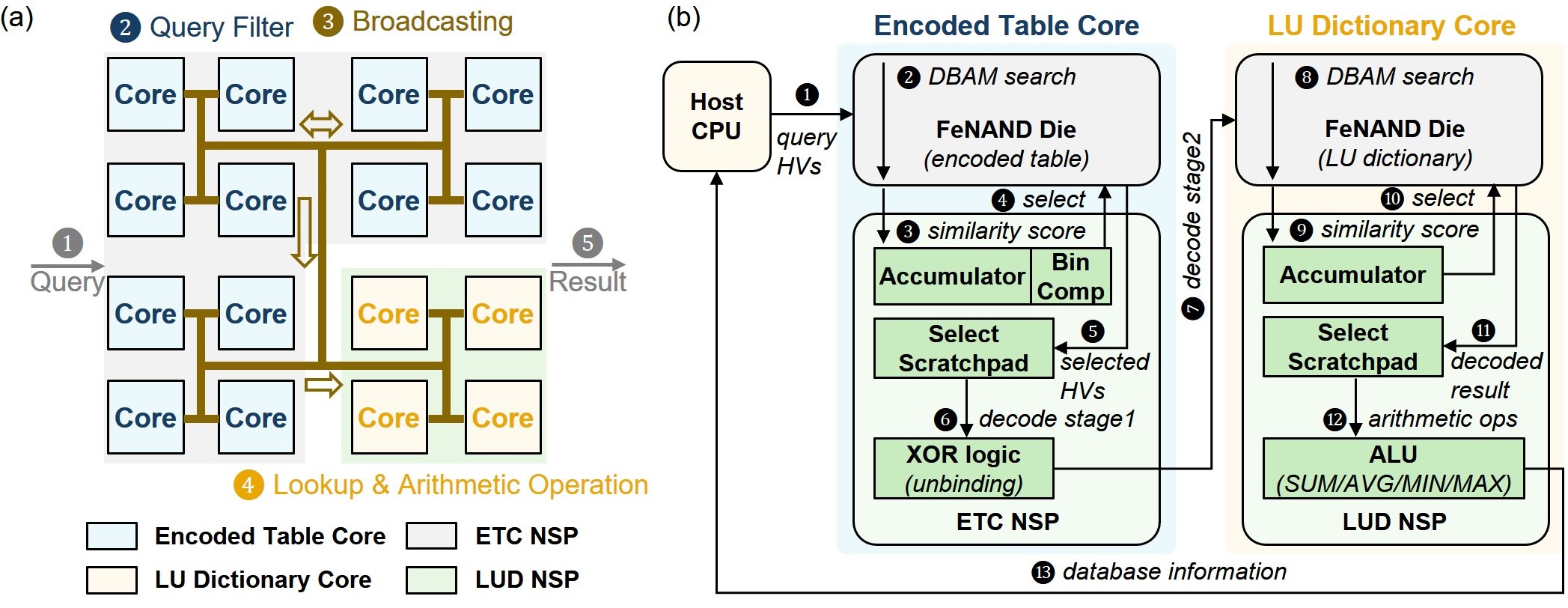}
    \caption{Overview of HDDB/ISP  (a) Reconfigurable H-tree connected architecture (b) Predicate evaluation dataflow}
    \label{fig:overview}
\end{figure}

The HDDB ISP acceleration system is a hardware-software co-design utilizing 3D heterogeneous integration to fuse 3D FeNAND storage tiles with specialized NSPs to minimize data movement. Driven by the high-throughput demands of hyperdimensional vector scans, this vertical stacking circumvents external I/O bottlenecks by exposing the massive internal bandwidth of the memory array directly to the compute logic. This interconnect links dedicated encoded table and lookup (LU) dictionary cores to their corresponding shared NSPs (ETC NSP and LUD NSP). The reconfigurable system-level architecture, illustrated in Figure \ref{fig:overview}(a), uses an H-tree network~\cite{7551379,song20251131}, managing query broadcasting and data routing.  The table cores store HDC-encoded database tables, while the dictionary cores hold LU dictionaries for decoding. The capacity ratio between two cores is configurable, as dictionary cores are often much smaller than table cores. Section \ref{ref:peripherals_design} details the specific circuit-level implementation of heterogeneous peripherals.

The predicate evaluation dataflow, detailed in Figure \ref{fig:overview}(b), begins at the host CPU. The host sends the query HVs and encoding plan to the encoded table cores (Step \encircle{1}). These cores execute a high-throughput in-situ DBAM search~\cite{pinge2025fenoms} against the stored encoded table HVs (Step \encircle{2}): For string predicates, the query HV is directly compared against the target column. Numerical predicate search is more complex, leveraging the 4-level (100 bins/level) encoding. An NSP accumulator processes the resulting similarity scores (Step \encircle{3}). For strings, this score directly identifies matching rows (Step \encircle{4}). For numerical queries, the scores are used to find the bin index of each cell; these indices are then compared externally to the query's target bin indices to select the correct rows (Step \encircle{4}). The selected cell HVs are buffered in a select scratchpad (Step \encircle{5}). An NSP parallel XOR logic unit then performs the decode stage 1, the HDC unbinding operation, on these selected HVs (Step \encircle{6}). The unbound HVs are forwarded to the LU dictionary core for the decode stage 2 (Step \encircle{7}). This core executes another DBAM search (Step \encircle{8}), comparing the intermediate unbound HVs against the LU dictionary. A second NSP accumulator processes these similarity scores (Step \encircle{9}) to select the final decoded results (Step \encircle{10}), storing them in the LU core's scratchpad (Step \encircle{11}). If aggregation is required, an NSP ALU performs arithmetic operations on these results (Step \encircle{12}). The final database information returns to the host CPU (Step \encircle{13}).

\subsection{In-situ Approximate Matching and Peripherals Design}
\label{ref:peripherals_design}

\begin{figure}[t]
    \centering
    \includegraphics[width=0.85\linewidth]{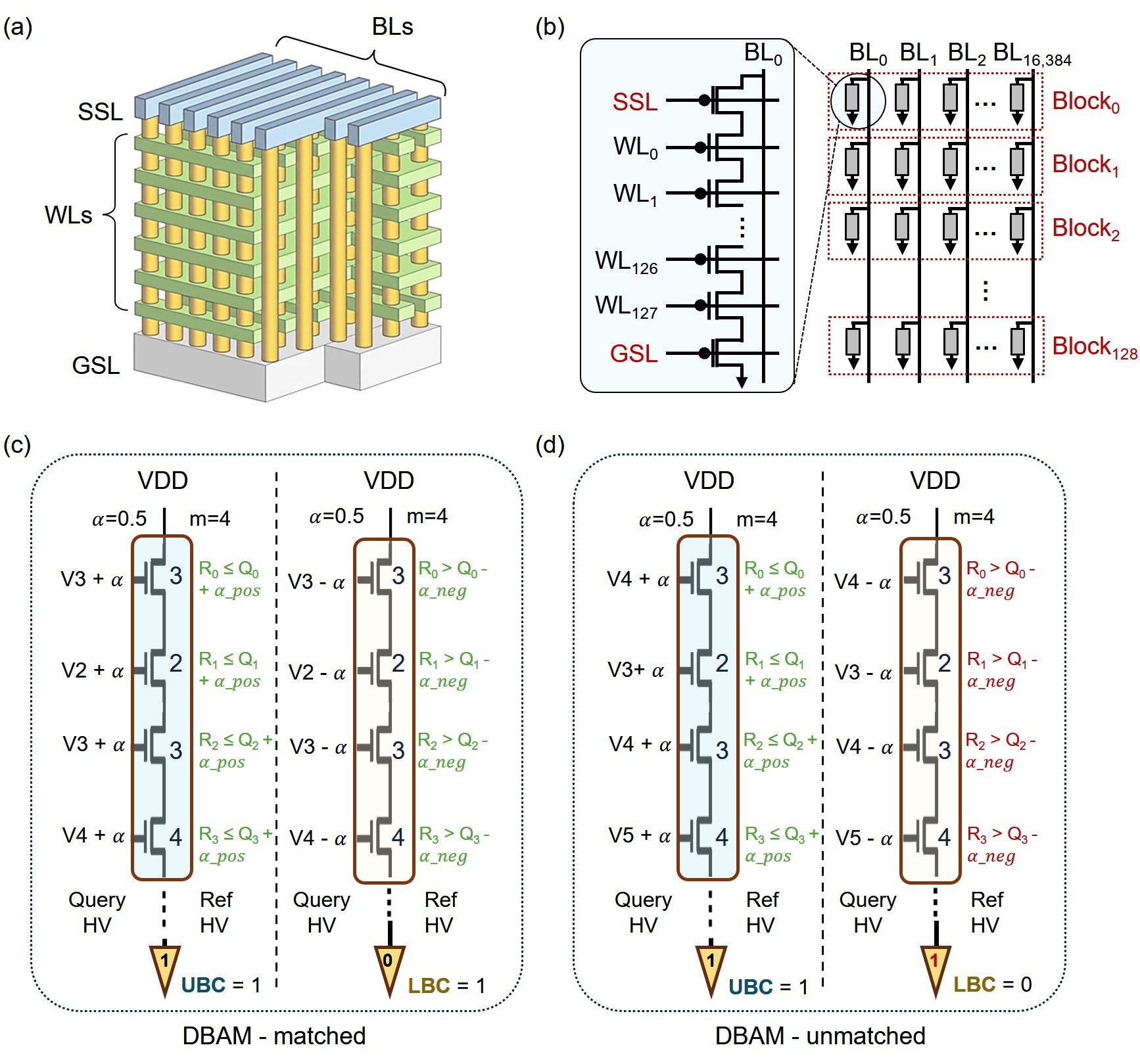}
    \caption{FeNAND-based Dual Boundary Approximate Matching (DBAM) (a) 3D FeNAND array structure (b) 3D FeNAND planes (c) DBAM matched case (d) DBAM unmatched case }
    \label{fig:DBAM}
\end{figure}

\textbf{In-situ Approximate Matching:} The fundamental operation for predicate evaluation in HDDB is the HDC similarity search (filtering). Unlike crossbar-based PIM architectures (e.g., ReRAM, PCM) that leverage parallel analog current summation, 3D NAND employs serial string connectivity that precludes standard matrix accumulation. To enable high-throughput search despite this structural constraint, we utilize the DBAM technique~\cite{pinge2025fenoms}. As shown in Figure~\ref{fig:DBAM}(a), the FeNAND tile is organized into multiple blocks, and bitlines (BLs) are shared across blocks. DBAM is engineered to exploit the serial connectivity of 3D NAND strings (Figure \ref{fig:DBAM}(b)) and maximize storage density by packing HVs into FeNAND Triple-Level Cells (TLC). Specifically, we map every three encoded bits to one cell (8 levels) using a fixed bijection Gray code. This encoding guarantees that threshold-voltage shifts between adjacent levels, a common issue in FeNAND, cause at most one bit to change. It thus preserves the similarity needed for reliable in-storage computation. The system reads these packed values by activating $k$ wordlines (WLs) simultaneously, comparing $k$ stored reference elements ($r_i$) to $k$ query elements ($q_i$) in a single operation. A configurable tolerance margin ($\alpha = 0.5$) provides robustness against $V_{TH}$ shifts and other device non-idealities.

The search operation completes in exactly two sensing cycles. The Upper Bound Check (UBC) applies a WL voltage of $q_i + \alpha_{\text{pos}}$. Due to the serial string, current flows only if all $k$ cells are below this bound, implementing a parallel AND operation.
\begin{equation}
\label{eq:2}
\text{UBC}_j = \prod_{i=kj}^{kj+k-1} \left[ r_i \leq q_i + \alpha_{\text{pos}} \right]
\end{equation}
The Lower Bound Check (LBC) applies a WL voltage of $q_i - \alpha_{\text{neg}}$ and blocks current if any cell exceeds the lower threshold.
\begin{equation}
\label{eq:3}
\text{LBC}_j = 1 - \prod_{i=kj}^{kj+k-1} \left[ r_i < q_i - \alpha_{\text{neg}} \right]
\end{equation}

As shown in Figure \ref{fig:DBAM}(c-d), these checks yield binary results per $k$-subset, which are aggregated by the NSP into a final similarity metric: $\text{Score} = \sum_j (\text{UBC}_j + \text{LBC}_j)$. This approach with $k$-way parallelism replaces conventional iterative MLC sensing~\cite{hyperoms,pinge2024rapidoms} with negligible hardware overhead. As demonstrated in FeNOMS~\cite{pinge2025fenoms} analysis, we select $k=8$ to achieve an optimal balance between parallel throughput and search accuracy. To support this array-level parallelism and process the resulting high-speed data streams, we implement a dedicated heterogeneous peripheral architecture.

\begin{figure}[t]
    \centering
    \includegraphics[width=1\linewidth]{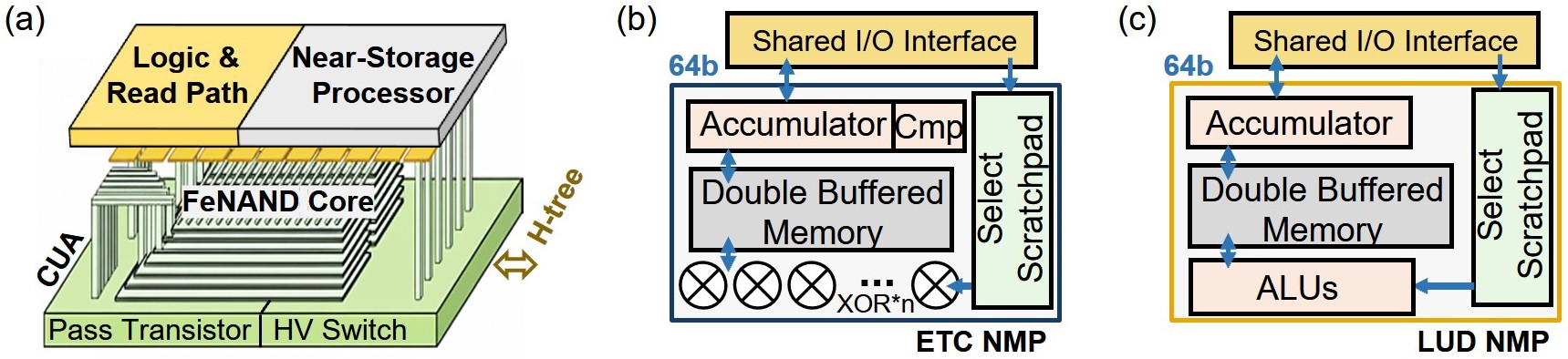}
    \caption{Design of (a) Heterogeneous integration (b) Encoded table near-storage processor (NSP) (c) LU dictionary NSP}
    \label{fig:peripherals}
\end{figure}

\textbf{Peripheral Design:} The HDDB/ISP peripheral design is a heterogeneous and reconfigurable architecture. The storage cores, shown in Figure \ref{fig:peripherals}(a), use 3D heterogeneous integration to stack the storage array atop a 65nm CMOS Under Array (CUA) for high-voltage management while placing performance-critical logic in a separate 7nm advanced logic layer. This upper layer integrates high-speed read path circuits including BL decoders and sense amplifiers alongside a dedicated NSP for each tile. By co-locating the NSP with the read circuitry, the design minimizes interconnect latency and maximizes internal bandwidth utilization. Each core is connected by the H-tree interconnect, providing low-latency, high-bandwidth data transfer between storage and processing.

While every core possesses an NSP, the specific logic configuration adapts to the core type to optimize area efficiency. The ETC NSP, detailed in Figure \ref{fig:peripherals}(b), executes the first stage of processing. It integrates a 7-bit bin index comparator to finalize numerical predicate selection by comparing the 4-level bin indices. It also contains a parallel bitwise XOR array for the HDC unbinding operation. The double-buffered memory structure concurrently supports intermediate result accumulation and output transfers. The LUD NSP, shown in Figure \ref{fig:peripherals}(c), contains a dedicated ALU that supports parallel SUM, AVG, MIN, and MAX operations for final aggregation. This decoupled architecture performs massive filtering in-situ. The filtered data is then processed in the NSPs for decoding and aggregation, enabling in-storage SQL analytics with massive parallelism and minimal data movement.

\subsection{HDDB In-Storage Mapping and Scheduling}

\begin{figure}[t]
    \centering
    \includegraphics[width=1\linewidth]{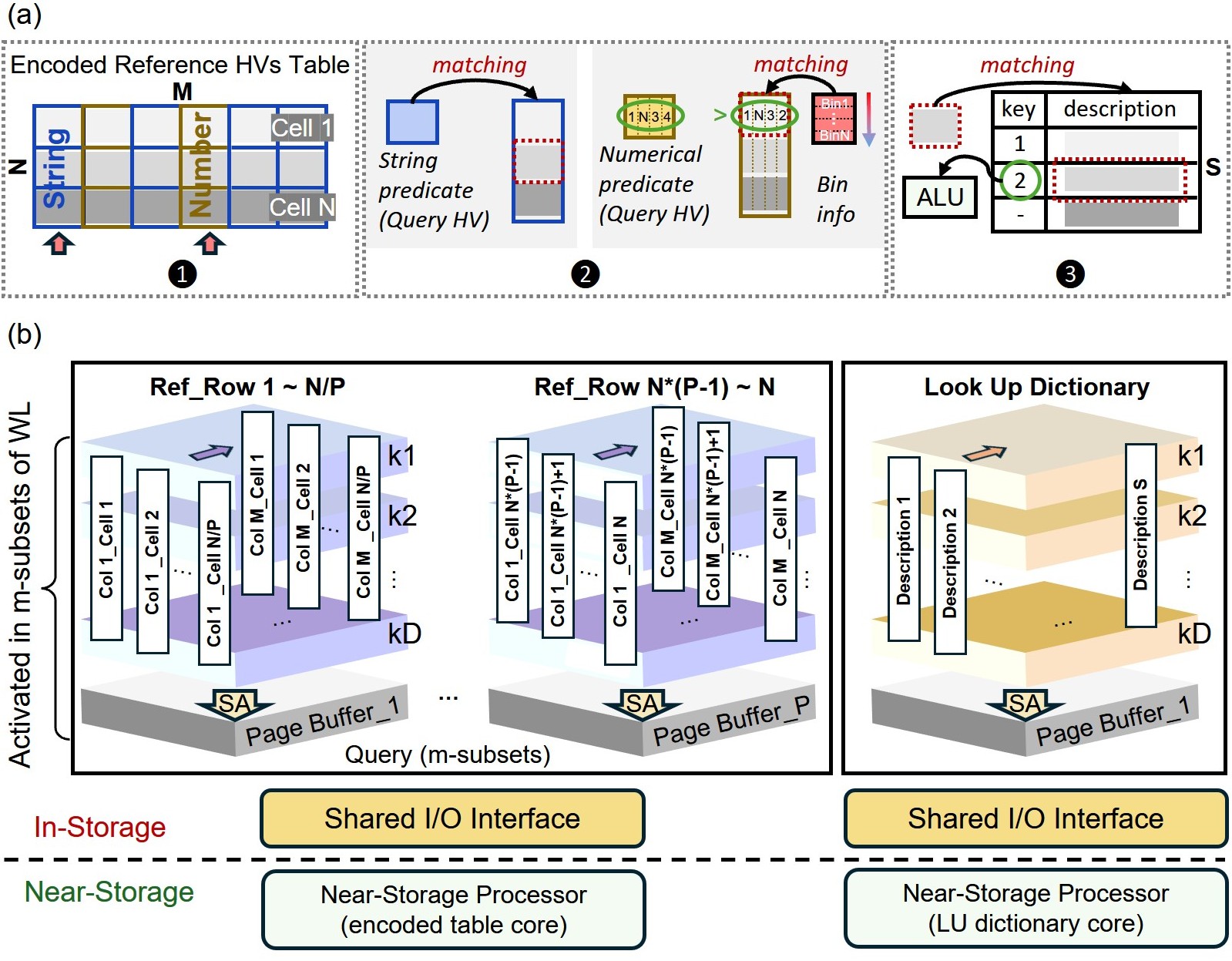}
    \caption{Mapping and scheduling (a) String and numerical cases (b) Data layout for encoded table and LU dictionary}
    \label{fig:layout}
\end{figure}

Figure \ref{fig:layout}(a) illustrates the multi-stage predicate evaluation algorithm. An HDDB subtable comprises M columns and N rows. The number of rows N scales with the Scale Factor (SF), while the column count M remains constant. The flow begins by selecting the target column based on a predetermined encoding plan. This plan serves as a schema map, specifying whether each column utilizes string or numeric encoding and delineating the dimension boundaries within the row vector. The in-storage DBAM engine then performs a parallel similarity search. This search adapts to the predicate type: string predicates involve direct HV comparison, whereas numerical predicates use the multi-level query HV to determine the correct bin index for each cell. Selected HVs are then unbound and used in a second DBAM search against the LU dictionary. This final lookup retrieves the decoded key values.

The HDDB system maps offline encoded tables onto the FeNAND array using a column-wise layout, as Figure \ref{fig:layout}(b) illustrates. Each cell of a given column is mapped onto a vertical FeNAND string. The data for a full column, containing up to 1.5 million rows (for an SF = 1 table), is evenly distributed across multiple planes. This layout achieves high plane-level parallelism and scales easily with SF changes. LU dictionaries are stored separately in dedicated LU dictionary core using this same mapping strategy. This workflow utilizes two specialized NSPs. The ETC NSP accumulates the initial scores and performs the HDC XOR unbinding operation. The LUD NSP handles the final dictionary search and executes aggregation functions such as SUM, AVG, MIN, or MAX within.

\section{Evaluation}
Because large fact tables dominate memory footprint and scan cost in SQL workloads, we evaluate HDDB on TPC-DS fact table. In Section~\ref{alg eval}, we generate random string and numerical predicates to test that HDDB can perform predicate evaluation reliably and decode information accurately, even in the presence of substantial hardware noise. In Section~\ref{system eval}, using these predicates, we construct two families of queries: (i) pure filter queries and (ii) filter+aggregation queries. For each family, we instantiate 1000 queries as common inputs, and use them to compare HDDB against SOTA database engines in terms of per query latency and energy consumption.

\subsection{HDDB Algorithm Evaluation}
\label{alg eval}

\begin{figure}[t]
    \centering
    \includegraphics[width=1\linewidth]{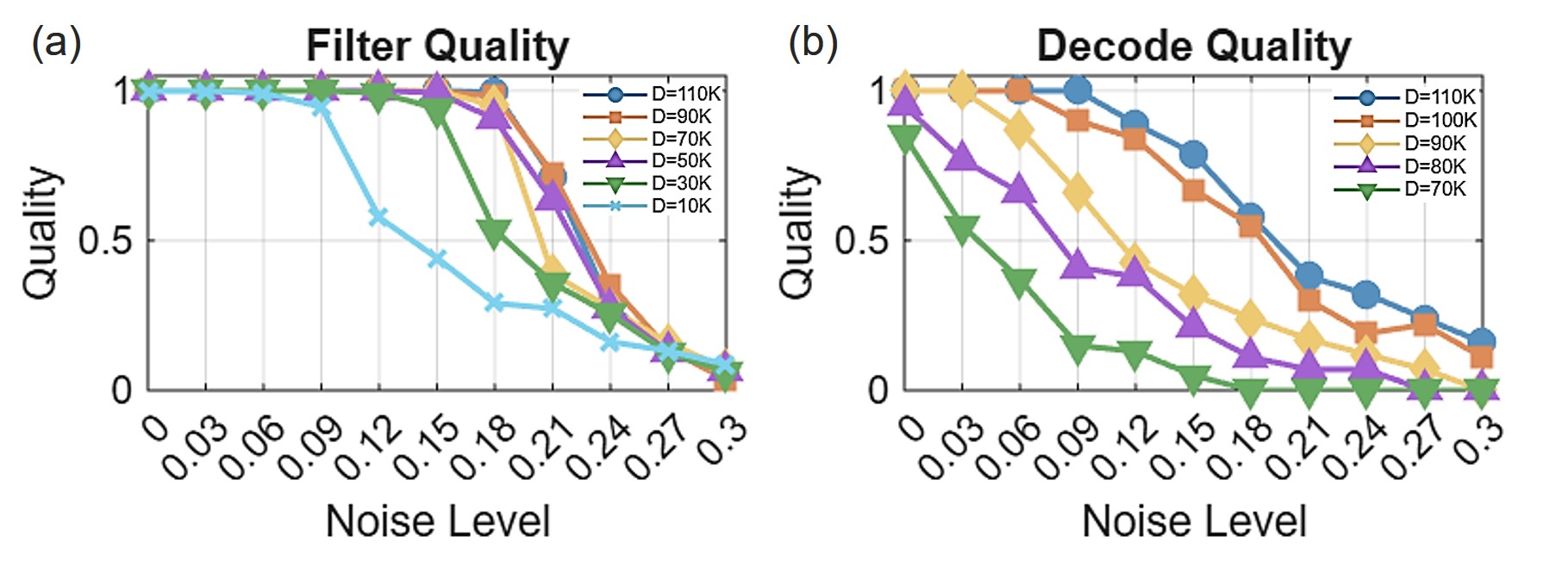}
    \caption{HDDB (a) Predicate evaluation (b) Decode quality}
    \label{fig:HDDBalg}
\end{figure}

We first evaluate the HDDB encoding and predicate algorithms in isolation. For encoding numerical data, we use an encoder with $m = 100$ bins per level and $n = 4$ recursive levels (effectively $100^4$ finest bins). To study how representation size affects both predicate correctness and decode accuracy, we sweep the per-row dimensionality between 70k and 110k bits.
To simulate device noise, we map binary bits to TLC cells and, for a given noise level, randomly choose that fraction of cells and shift their programmed state by $\pm 1$ level, reflecting the fact that most retention and sense-margin failures in MLC NAND appear as adjacent-level shifts. This model fits comfortably with prior work\cite{pinge2025fenoms}, which models cell-level non-idealities as Gaussian threshold-voltage noise, $V_T \sim \mathcal{N}(0,0.2^2)\,\text{V}$ within a 6.5\,V window; the resulting $V_T$ perturbations effectively induce these $\pm 1$-level errors near decision boundaries.
For each row dimensionality, we encode the TPC-DS fact table into hypervectors, generate random string and numerical predicates, and run HDDB’s predicate evaluation algorithm to determine which rows satisfy each predicate. For every predicate, we measure quality as (i) whether the predicate outcome matches exactly the ground-truth SQL execution and (ii) whether we can correctly decode the original contents of the selected rows from their hypervectors. Figure~\ref{fig:HDDBalg}(a) shows that under 100K to 110K dimensionality, HDDB’s predicate evaluation remains perfectly accurate up to a 0.15 noise level. In contrast, Figure~\ref{fig:HDDBalg}(b) shows that decoding the original cell information is more sensitive to noise, but still able to achieve perfect decoding at a 0.1 noise level. This shows that increasing dimensionality improves HDDB’s ability to process predicates correctly and to reliably recover the original data for any downstream analytics.

\subsection{HDDB In-Storage Accelerator Evaluation}
\label{system eval}

\subsubsection{Experimental Setup}

\noindent\textbf{Baselines:} We compare our HDDB against widely used SQL engines: DuckDB~\cite{raasveldt2019duckdb}, PostgreSQL~\cite{stonebraker1986design}, HeavyDB~\cite{root2016mapd,heavydb_docs}, and a Dask-cuDF backed SQL engine~\cite{daskcudf,dasksql}. These represent SOTA or popular choices for SQL workloads on commodity platforms. Baselines were executed on platforms featuring the Intel Ultra7 155H/DDR5-5600 (16GB) and the NVIDIA RTX 4070/GDDR6 (8GB). To better contextualize HDDB relative to hardware database accelerators beyond CPU/GPU, we also report comparisons with recent accelerator-based systems, including SQL2FPGA~\cite{lu2024sql2fpga}, Darwin~\cite{kim2024darwin}, and pimDB~\cite{bernhardt2023pimdb}. For system evaluation in this section, we use a per-row dimension of 110K.


\noindent\textbf{Hardware Configurations:} We model the 3D FeNAND core based on the 3D NAND architecture~\cite{ISP, SLC3DNAND} with 128 WLs, 16,384 BLs, and 128 blocks, assuming a FeNAND z-scaling ratio of $k=4$. This scaling ratio $k=4$, derived from in-house modeling and inference from studies on ferroelectric material properties \cite{das2023ferroelectric,venkatesan2024disturb,kim2021ferroelectric}, is applied for the  RC delay modeling and analysis to compare with conventional 3D NAND. System-level parameters are also detailed in Table \ref{table:hardware_params}.

\begin{table}[htbp]
\centering
\caption{Hardware configurations of HDDB/ISP accelerator}
\renewcommand{\arraystretch}{1.1}
\setlength{\tabcolsep}{2pt}
\scriptsize

\resizebox{\columnwidth}{!}{%
\begin{tabular}{|l|l|l|l|}
\hline
\multicolumn{4}{|c|}{\textbf{3D FeNAND Core Parameters}} \\
\hline
Parameter & \multicolumn{3}{l|}{WL = 128, BL = 16,384, \#Blocks = 128, \#Tiles = 1, 1 NSP/core} \\
\hline
SSL, WL, BL Pitch & 220 nm, 500 nm, 100 nm & Capacity & 3GB per TLC Core \\
\hline
GSL, SSL, Blocks & 16, 16, 128 & FeNAND z-scaling & \(k = 4\) \\
\hline
Plane area (mm\(^2\)) & 0.738198 & WL, SSL, BL Read & 1 V, 4.5 V, 0.2 V \\
\hline
Read Latency & 50–100 µs/page read & Read Energy & 2.28 pJ/bit\\
\hline
Write Latency & 0.2–0.8 ms/page program & Write Energy & 34.2 pJ/bit \\
\hline
\multicolumn{4}{|c|}{\textbf{3D FeNAND System Parameters}} \\
\hline
Organization &  1 FeNAND flash controller/2TB & Interconnect & 4 core per H-trees \\

\hline
\multicolumn{4}{|c|}{\textbf{NSP (Encoded table core)}} \\
\hline
XOR logic &  42 parallel bitwise XOR logic & Select Scratchpad & 20 KB \\
\hline
Double buffered SRAM & 2 KB (same as LUD NSP)  & 7-bit Bin Cmp. & 5-parallel comparator \\
\hline
\multicolumn{4}{|c|}{\textbf{NSP (LU dictionary core)}} \\
\hline
ALU & 2 units (ADD/AVG/MIN/MAX) & Select Scratchpad & 20 KB \\
\hline
\end{tabular}}
\label{table:hardware_params}
\end{table}

\noindent\textbf{Evaluation Platform:} We develop an in-house simulator for the HDDB/ISP system. We model the 3D FeNAND arrays and logic \& read path circuits in SystemVerilog to ensure device-level accuracy, and we synthesize the Register Transfer Level (RTL) design using Synopsys Design Compiler with a 64\,nm CMOS PDK (1 V) and a 7nm CMOS advanced PDK (0.7V) at 1\,GHz.

\subsubsection{Comparison with SOTA Database Engine}
\begin{figure}[t]
    \centering
    \includegraphics[width=0.9\linewidth]{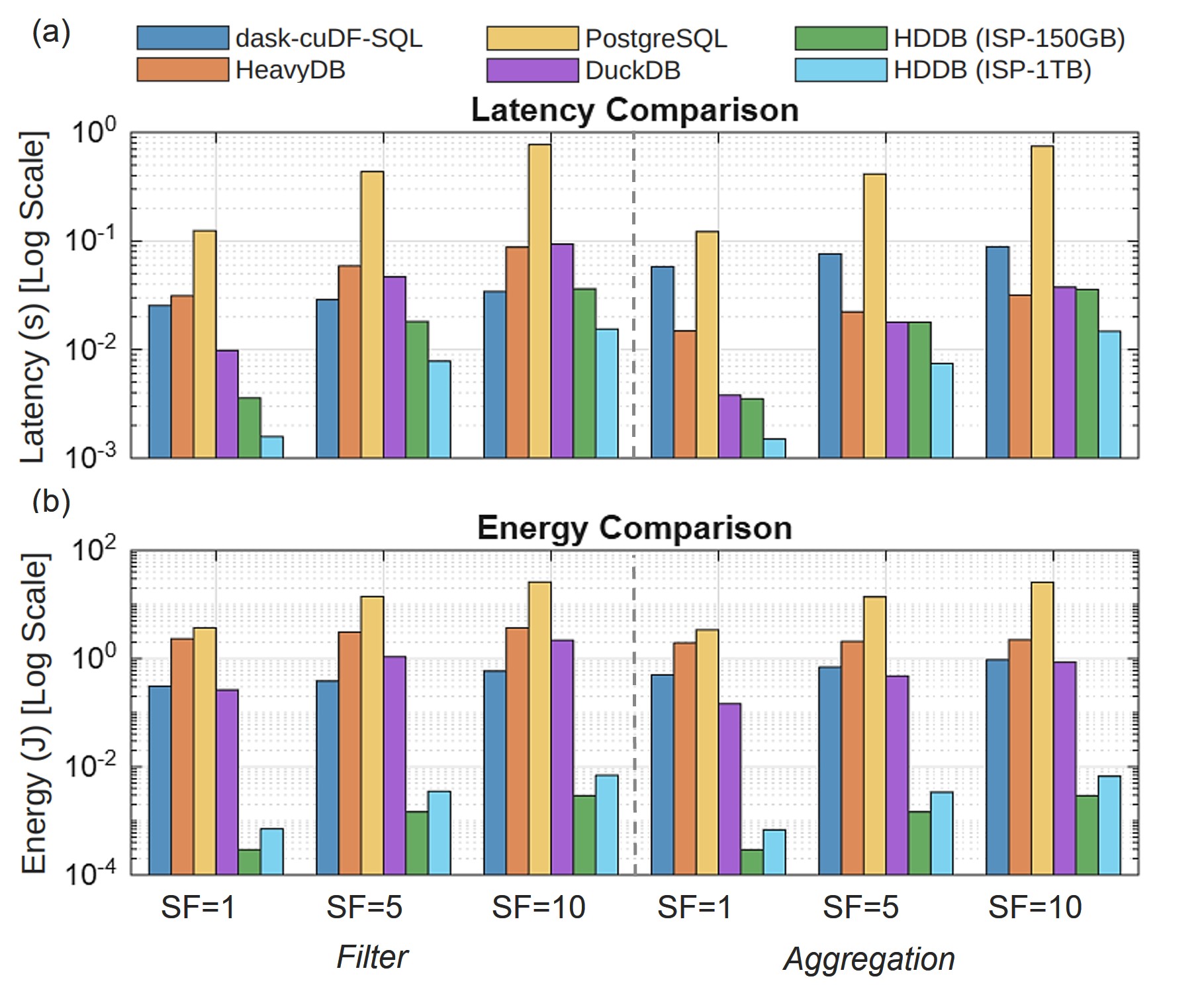}
    \caption{Performance comparison over SOTA database engine on average (a) Latency (b) Energy consumption}
    \label{fig:HDDBhw}
\end{figure}

We evaluate HDDB against two configurations (ISP-150GB and ISP-1TB) and four SOTA database engines. These include two GPU-accelerated systems (dask-cuDF-SQL~\cite{daskcudf,dasksql}, HeavyDB~\cite{root2016mapd,heavydb_docs}) and two high-performance CPU-based systems (PostgreSQL~\cite{stonebraker1986design}, DuckDB~\cite{raasveldt2019duckdb}). Figure \ref{fig:HDDBhw} shows the latency and energy consumption on a logarithmic scale across TPC-DS filter and aggregation workloads at SF=1, 5, and 10.

Figure \ref{fig:HDDBhw}(a) shows that both HDDB configurations consistently achieve the lowest latency across all workloads and scale factors. The HDDB (ISP-1TB) configuration achieves up to 78.7$\times$ speedup over PostgreSQL and 19.8$\times$ over HeavyDB for the SF=1 filter workload. The energy efficiency improvements shown in Figure \ref{fig:HDDBhw}(b) are more significant. HDDB consumes orders of magnitude less energy than any baseline. The HDDB (ISP-150GB) system is up to 12,636$\times$ more energy-efficient than PostgreSQL. Meanwhile, the 1TB configuration saves 3,258$\times$ the energy consumed by HeavyDB for SF=1 filter, demonstrating the benefits of eliminating data movement.

Table \ref{table:sota_comparison} further contextualizes HDDB against other hardware-accelerated database engines. While systems like SQL2FPGA, Darwin, and pimDB rely on FPGA co-processing or DRAM-based PIM, HDDB is a true ISP system on MLC FeNAND flash. This approach moves computation directly to the data, eliminating the I/O bottleneck rather than just reducing it. The resulting 80.6$\times$ speedup of our system is highly competitive with the performance gains reported by these other specialized hardware solutions.

\begin{table}[htbp]
\centering
\caption{Comparison of SOTA hardware-accelerated database engines (SF = 1)}
\label{table:sota_comparison}
\scriptsize 
\renewcommand{\arraystretch}{1.1} 
\setlength{\tabcolsep}{1pt} 

\begin{tabularx}{\columnwidth}{|l|
    >{\raggedright\arraybackslash}X|
    >{\raggedright\arraybackslash}X|
    >{\raggedright\arraybackslash}X|
    >{\raggedright\arraybackslash}X|}
\hline
\textbf{Feature} & \textbf{HDDB (Ours)} & \textbf{SQL2FPGA~\cite{lu2024sql2fpga}} & \textbf{Darwin~\cite{kim2024darwin}} & \textbf{pimDB~\cite{bernhardt2023pimdb}} \\
\hline
Hardware & MLC Flash-based ISP & CPU-FPGA (Intel Xeon-Xilinx Alveo) & GDDR6 DRAM PIM & UPMEM DRAM PIM \\
\hline
\multicolumn{5}{|c|}{\textbf{Evaluation \& Baselines}} \\
\hline
Datasets & \textbf{TPC-DS (Filter/Agg.)} & TPC-H, TPC-DS (Q1-5) & TPC-H, Basic Query Operators & TPC-H (Scan/Select) \\
\hline
Baselines & \textbf{SOTA:} See Figure~\ref{fig:HDDBhw} & Big Data Frmwk: Apache Spark & Trad.\ Impl.: Optimized C++ & Trad.\ Impl.: Optimized C++ \\
\hline
\multicolumn{5}{|c|}{\textbf{Performance \& Efficiency}} \\
\hline
Latency Speedup & Up to \textbf{80.6$\times$ } & $\sim$10.1$\times$ & 4$\times$--44$\times$  & Up to 68$\times$ \\
\hline
Energy Efficiency & \textbf{88$\times$--12.6k$\times$} & $\sim$9.2$\times$ & $\sim$7$\times$ & Not Reported \\
\hline
\end{tabularx}
\end{table}

\subsubsection{Area and Power Analysis}

\noindent\textbf{Area and Power Breakdown:} Table \ref{tab:HDDBbreakdown} presents the post-synthesis area and power breakdown for the ETC and LUD NSPs synthesized at 7nm, 1GHz, 0.7V. On-chip storage components, specifically the select scratchpad and double buffered SRAM, occupy $\sim$80\% of the total footprint in both processors. The specialized compute units remain compact: The 7-bit bin comparator and parallel XOR array in the ETC NSP together utilize less than 0.6\% of the area while the ALUs in the LUD NSP require less than 1.9\%. This demonstrates that offloading predicate logic to near-storage processors incurs minimal hardware overhead.

\noindent\textbf{Overhead Analysis:} System-level analysis further validates that the logic fits easily within the system footprint: a single 3D FeNAND storage tile consumes 1.2964 mW and occupies 0.738 mm$^2$~\cite{pinge2025fenoms}, whereas the NSP occupies only $\sim$0.013 mm$^2$. Using the storage tile as a baseline, the NSP logic incurs a vertical area overhead of just 1.8\% relative to the tile size. This indicates that the logic is densely integrated within the vertical shadow of the storage array without expanding the total chip footprint and ensures zero loss in FeNAND memory capacity. At the package level, the SM2508 controller adds a fixed power overhead of 3.5 W within a 225 mm$^2$ footprint; the scalable design results in a total system area of 75.1 mm$^2$ for a 150GB system and 512.6 mm$^2$ for a 1TB implementation.

\begin{table}[htbp]
\centering
\caption{Area and Power Breakdown of ETC NSP \& LUD NSP }
\label{tab:HDDBbreakdown}
\scriptsize 
\renewcommand{\arraystretch}{1.1} 
\setlength{\tabcolsep}{2pt} 
\begin{tabularx}{\columnwidth}{|l|
    >{\raggedright\arraybackslash}X|
    >{\raggedright\arraybackslash}X|
    >{\raggedright\arraybackslash}X|
    >{\raggedright\arraybackslash}X|}
\hline
\multirow{2}{*}{\textbf{Component}} & \multicolumn{2}{c|}{\textbf{ETC NSP Logic}} & \multicolumn{2}{c|}{\textbf{LUD NSP Logic}} \\
\cline{2-5}
 & \textbf{Area ($\mu$m$^2$)} & \textbf{Power (mW)} & \textbf{Area ($\mu$m$^2$)} & \textbf{Power (mW)} \\
\hline
Accumulator & 2520 (19.25\%) & 18.5 (46.49\%) & 2520 (18.99\%) & 18.5 (43.10\%)  \\
\hline
Select Scratchpad & 8043 (61.45\%) & 15 (37.70\%) & 8043 (60.62\%) & 15 (34.95\%) \\
\hline
Double buffered SRAM & 2457 (18.77\%) & 5 (12.57\%) & 2457 (18.52\%) & 5 (11.65\%) \\
\hline
7-bit Bin Comparator & 20.37 (0.16\%) & 0.412 (1.04\%) &  -- & -- \\
\hline
Parallel XOR Array & 49.31 (0.38\%) & 0.879 (2.21\%) &  -- & -- \\
\hline
ALUs & -- & -- & 248.53 (1.87\%) & 4.424 (10.31\%) \\
\hline
\textbf{Total} & \textbf{13089.68} & \textbf{39.791} & \textbf{13268.53} & \textbf{42.924} \\
\hline
\end{tabularx}
\end{table}


\section{Conclusion}
In this paper, we presented HDDB, a hardware–software co-design that, to our knowledge, is the first system to apply Hyperdimensional Computing (HDC) to SQL databases and to execute large-scale predicate evaluation directly inside noisy FeNAND storage. HDDB merges HDC’s massively parallel, noise-tolerant computation model with FeNAND’s ultra-high density and in-storage compute capability. Our evaluation on TPC-DS fact tables shows that HDDB preserves correct predicate outcomes under substantial device noise while achieving up to 80.6× lower latency and 12,636× lower energy consumption than CPU/GPU SQL engines. These results indicate HDDB offers an efficient, noise-robust substrate for memory-centric SQL workloads. More broadly, HDDB points to a new class of database accelerators that co-design data representations, query algorithms, and emerging memories, and can be extended to richer SQL operators and queries.

\section{ACKNOWLEDGMENT}
This work was supported in part by PRISM and CoCoSys, centers in JUMP 2.0, an SRC program sponsored by DARPA (SRC grant number - 2023-JU-3135). This work was also supported by NSF grants \#2003279, \#1911095, \#2112167, \#2052809, \#2112665, \#2120019, \#2211386.


\bibliographystyle{ACM-Reference-Format}
\bibliography{Reference}

\end{document}